\documentclass[aps,prd,superscriptaddress,nofootinbib,showpacs]{revtex4}
\usepackage{graphicx}
\usepackage{epsfig}
\usepackage{color}
\usepackage{amsmath}
\usepackage{amssymb}
\begin{document}
\title{Naked Singularity Formation In $f({\mathcal R})$ Gravity}
\author{A. H. Ziaie}
\email{ah.ziaie@gmail.com} \affiliation{Department of
Physics, Shahid Beheshti University, Evin, 19839 Tehran, Iran}
\author{K. Atazadeh}
\email{k-atazadeh@sbu.ac.ir} \affiliation{Department of Physics, Azarbaijan University of Tarbiat Moallem, 53714-161 Tabriz, Iran}
\author{S. M. M. Rasouli}
\email{m-rasouli@sbu.ac.ir} \affiliation{Department of Physics,
Shahid Beheshti University, Evin, 19839 Tehran, Iran}
\date{\today}
\begin{abstract}
We study the gravitational collapse of a star with barotropic equation
of state $p=w\rho$ in the context of $f({\mathcal R})$ theories of
gravity. Utilizing the metric formalism, we rewrite the field
equations as those of Brans-Dicke theory with vanishing coupling
parameter. By choosing the functionality of Ricci scalar as
$f({\mathcal R})=\alpha{\mathcal R}^{m}$, we show that for an
appropriate initial value of the energy density, if $\alpha$ and $m$
satisfy certain conditions, the resulting singularity would be naked,
violating the cosmic censorship
conjecture. These conditions are the ratio of the mass function to the
area radius of the collapsing ball, negativity of the effective
pressure, and the time behavior of the Kretschmann scalar. Also, as long as
parameter $\alpha$ obeys certain conditions, the
satisfaction of the weak energy condition is guaranteed by the
collapsing configuration.
\end{abstract}
\maketitle
\section{Introduction}
\textbf{E}instein's General theory of relativity is the classical
theory of one of the four fundamental forces, gravity, which is
the weakest but most dominant force of nature governing phenomena
at large scales, and is described by a mathematically well-founded
and elegant structure i.e., differential geometry of curved
spacetime. The Einstein's field equations, a system of non-linear
partial differential equations, relate the geometric property of
spacetime to the four-momentum (energy density and linear
momentum) of matter fields leading to precise predictions that
have received considerable experimental confirmations with high
accuracy such as solar system tests (see~\cite{Solar} and
references therein).~One of the most engrossing but open debates
in general relativity is that of the final fate of gravitational
collapse with possibility of the existence of spacetime
singularities, the ultra-strong gravity regions where the
densities and spacetime curvatures blow up, leading to a spacetime
which is geodesically incomplete~\cite{Hawking-Ellis} and the
structure of any classical theory of fields is vanquished. A star
with a mass many times than that of the Sun would undergo a
continual gravitational collapse due to its self-gravity without
achieving an equilibrium state in contrast to a neutron star or a
white dwarf. Then, according to singularity theorems established
by Hawking and Penrose~\cite{Hawking-Penrose} a singularity is
reached as the collapse endstate. Such a singularity may be a
black hole hidden from external observers by an event horizon or
visible to the outside Universe (naked singularity). In the latter
collapse procedure, the information on super-dense regions can be
transported via suitable non-spacelike trajectories to a distant
observer. Although the occurrence of a spacetime singularity as
the final outcome of a collapse scenario is proved by the
singularity theorems, they do not specify the nature of such a
singularity. The cosmic censorship conjecture first articulated by
Penrose~\cite{CCC} states that a black hole is always formed in
complete gravitational collapse of reasonable matter fields, or a
physically reasonable spacetime contains no naked singularities.
However, up to now many exact solutions of Einstein's field
equations describing singularities, not hidden behind an event
horizon of spacetime, are known. A remarkable study is the one by
Shapiro and Teukolsky, who showed numerically that gravitational
collapse of a spheroidal dust may end in a naked singularity
~\cite{Shapiro}. Also many exact solutions of Einstein's field
equations with a variety of field-sources admitting naked
singularities have been surveyed. The examples studied so far
include gravitational collapse of a pressure-less
matter~\cite{Dust-Patil}, radiation~\cite{Radiation-Patil},
perfect fluids~\cite{PF-Patil}, imperfect fluids~\cite{IPF-Patil}
and null strange quark fluids~\cite{NSQF}. Beside general theory
of relativity, there exist alternative theories of gravity
explaining gravitational phenomena. Such theories have been
studied for a long time~\cite{Alternative}. From the theoretical
standpoint there has been many attempts to correct the
Einstein-Hilbert action in order to renormalize general relativity
to build a quantum theory of gravity or at least some effective
action~(including the low-energy limit of string theories), or to
quantize the scalar fields in curved spacetimes~\cite{Callan}.
From the observational point of view, the discovery of current
acceleration of the Universe using CMB \textbf{I}a
supernova~\cite{CMB} suggests that such acceleration may be
explained within the framework of general relativity by assuming
that $76\%$ of energy content of the Universe is filled with a
mysterious form of $dark~energy$ with equation of state
$p\sim-\rho$ (where $\rho$ and $p$ are the energy density and
pressure of the cosmic fluid, respectively). Another possibility
is to include a cosmological constant $\Lambda$ of a very small
magnitude in Einstein's field equation, but encounters such
difficulties as the well-known cosmological constant problem and
the coincidence problem. Another alteration is $f({\mathcal R})$
theories of gravity~\cite{FR} where the Ricci scalar in the
Einstein-Hilbert Lagrangian is replaced by a general function of
it, providing alternative gravitational models for dark energy
since the explanation of the cosmic acceleration comes back to the
fact that we do not understand gravity at large scales. Such
theories can describe the transition from deceleration to
acceleration in the evolution of the
Universe~\cite{Nojiri-Odintsov}. Moreover, the coincidence problem
may be solved simply in such theories by the Universe expansion.
Also some models of modified theories of gravity are predicted by
string/M-theory considerations~\cite{MString}. Recently, it has
been shown that the accelerated expansion of the Universe may be
the result of a modification to the Einstein-Hilbert action in the
context of higher order gravity theories~\cite{Farhoudi}. Our
purpose here is to consider the gravitational collapse of a star
within the framework of $f({\mathcal R})$ theories of gravity,
whose matter content obeys the barotropic equation of state,
$p=w\rho$.
We investigate the conditions under which the resulting
singularity may be naked or not. In Section \textbf{II} we apply
the metric formalism to the action of $f({\mathcal R})$
gravity~\cite{Metric-Formalism} and rewrite it as that of the
Brans-Dicke theory with $\omega_{BD}=0$. Choosing $f({\mathcal
R})=\alpha{\mathcal R}^{m}$~\cite{Capozzillo} in Section
\textbf{III} and fixing the corresponding potential, we find $m$
as a function of initial energy density and $\alpha$. In Section
\textbf{IV} we study the behavior of the expansion parameter which
is the key factor in examining the formation or otherwise of
trapped surfaces during the dynamical evolution of the collapse
scenario. In Section \textbf{VI} we examine the global features of
the nakedness of the resulting singularity by investigating the
behavior of the Kretschmann scalar as a function of time. In order
to fully complete the model we utilize the Vaidya metric to match
the interior spacetime to that of the exterior one.

\section{$f({\mathcal R})$ Field equations}

 We begin by the general action in modified
theories of gravity given by
\begin{equation}\label{eq1}
{\mathcal A}=\frac{1}{2\kappa}\int d^{4}x\sqrt{-g}f({\mathcal R})+
{\mathcal A}_{matter}(g_{\mu\nu},\psi),
\end{equation}
where $\kappa=8\pi G$, G is the gravitational constant, g is the
determinant of the metric, ${\mathcal R}$ represents the Ricci
scalar and $\psi$ collectively denotes the matter fields.
Introducing an auxiliary field $\Psi$, one can write the
dynamically equivalent action as (See \cite{Faraoni} and
references therein)
\begin{equation}\label{eq2}
{\mathcal A}=\frac{1}{2\kappa}\int
d^{4}x\sqrt{-g}\left[f(\Psi)+f'(\Psi)({\mathcal R}-\Psi)\right]+
{\mathcal A}_{matter}(g_{\mu\nu},\psi),
\end{equation}
where variation with respect to $\Psi$ leads to the following
equation as
\begin{equation}\label{eq3}
f''(\Psi)({\mathcal R}-\Psi)=0.
\end{equation}
If $f''(\Psi)\neq0$, one can then recover action (\ref{eq1})
by setting $\Psi={\mathcal R}$. Redefining the field $\Psi$ by
$\phi=f'(\Psi)$ and setting
\begin{equation}\label{eq4}
V(\phi)=\Psi(\phi)\phi-f(\Psi(\phi)),
\end{equation}
action (\ref{eq2}) will take the following form
\begin{equation}\label{eq5}
{\mathcal A}=\frac{1}{2\kappa}\int
d^{4}x\sqrt{-g}\left[\phi{\mathcal R}-V(\phi)\right]+ {\mathcal
A}_{matter}(g_{\mu\nu},\psi),
\end{equation}
which corresponds to the Jordan frame representation of the action
of Brans-Dicke theory with Brans-Dicke parameter~$\omega_{BD}=0$.
 Brans-Dicke theory with $\omega_{BD}=0$ is sometimes called {\it
massive dilaton gravity}~\cite{Wands} which was originally
suggested in~\cite{O'Hanlon} in order to generate a Yukawa term in
the Newtonian limit. Extremizing the action yields the following
field equations as (we set $\kappa=8\pi G=1$ in the rest of
this~paper)
\begin{equation}\label{eq6}
G_{\mu \nu} = T^{({\rm eff})}_{\mu \nu},
\end{equation}
and
\begin{equation}\label{eq7}
3\Box\phi+2V(\phi)-\phi\frac{dV(\phi)}{d\phi}=T^{m},
\end{equation}
where $T^{m}$ stands for the trace of $T^{m}_{\mu \nu}$ and the
subscript ``$m$'' refers to the matter fields (fields other than
$\phi$) and we have defined the effective stress-energy tensor as
\begin{equation}\label{eq8}
T^{({\rm eff})}_{\mu \nu} = \frac{1}{\phi}\left(T^{\rm m}_{\mu
\nu}+ T^{\phi}_{\mu \nu}\right),
\end{equation}
with
\begin{equation}\label{eq9}
T^{\phi}_{\mu\nu}=
(\nabla_{\mu}\nabla_{\nu}\phi-g_{\mu\nu}\Box\phi)
-\frac{1}{2}g_{\mu\nu}V(\phi),
\end{equation}
and
\begin{equation}\label{eq10}
T^{\mu m}_{\nu} = {\rm
diag}\left(\rho_{m},p_{m},p_{m},p_{m}\right),
\end{equation}
being the stress-energy tensors of the scalar field and a perfect
fluid, respectively.

\section{Gravitational collapse of a homogeneous cloud with $f({\mathcal R})=\alpha{\mathcal R}^{m}$}

Let us now build and investigate a homogeneous class of collapsing
models in $f({\mathcal R})$ gravity with $m\neq0$, where the
trapping of light is avoided till the formation of singularity,
allowing the singularity to be visible to outside observers. In
order to achieve our purpose we examine a spherically symmetric
homogeneous scalar field, $\Phi=\Phi(\tau)$ originating from
geometry. Since the interior spacetime is a dynamical one, we
parameterize its line element as follows
\begin{equation}\label{eq11}
 ds^2=-d\tau^{2} + a^{2}(\tau)(dr^{2}+ r^{2}d\Omega^{2}),
\end{equation}
where $\tau$ is the proper time of a free falling observer whose
geodesic trajectories are distinguished by the comoving radial
coordinate $r$ and $d\Omega^{2}$ is the standard line element
on the unit $2$-sphere.  It is worth mentioning that here we
assume that starting from the homogeneous initial data, the collapsing
configuration remains homogeneous till the singularity is formed.
But as the collapse proceeds there may be some inhomogeneities
occurring throughout the collapse scenario, the existence of
which can be investigated by perturbation theory, that
is, imposing inhomogeneous perturbations on the energy density,
scale factor and BD scalar field and then see whether the
terms rising from inhomogeneity are dominant in the formation of
the singularity or not. Here we do not deal with such an issue but
for more details the reader may consult \cite{per} and references
there in. Since the presence of matter acting as a ``seed'' field
prompts the collapse of the BD scalar field, we have considered
perfect fluid models obeying barotropic equation of state as
\begin{equation}\label{eq12}
p_{m} = w \rho_{m}.
\end{equation}
Using the conservation equation for the matter
($\nabla^{\alpha}T^{m}_{\alpha\beta}=0$) together with the use of
above equation, one gets the following relations between $\rho_{m}
,~ p_{m}$ and the scale factor as
\begin{equation}\label{eq13}
\rho_{m} =  \rho_{_{_{0}}m}a^{-3(1+w)};~~p_{m} =
w\rho_{_{_{0}}m}a^{-3(1+w)},
\end{equation}
where $\rho_{_{_{0}}m} = \rho_{m}(a=1)$, is the initial value of
energy density of matter on the collapsing volume. Making use of
equation (\ref{eq8}) and equation (\ref{eq11}) one finds the
following equations for the effective stress-energy tensor as
\begin{equation}\label{eq14}
\rho_{_{({\rm eff})}} =-T^{\tau}\,_{\tau}^{({\rm eff})}=
\frac{1}{\phi}\left(
\rho_{m}+\rho_{_{\phi}}\right)=\frac{1}{\phi}\left[\rho_{m}-3\frac{\dot{a}}{a}\dot{\phi}
+\frac{V(\phi)}{2}\right],
\end{equation}
and
\begin{equation}\label{eq15}
p_{_{({\rm eff})}} =T^{r}\,_{r}^{({\rm
eff})}=T^{\theta}\,_{\theta}^{({\rm
eff})}=T^{\varphi}\,_{\varphi}^{({\rm eff})}=
\frac{1}{\phi}\left(p_{_{m}}
+p_{\phi}\right)=\frac{1}{\phi}\left[p_{m}+2\frac{\dot{a}}{a}\dot{\phi}+\ddot{\phi}-\frac{V(\phi)}{2}\right],
\end{equation}
with all other off-diagonal terms being zero and the radial and
tangential profiles of pressure are equal due to the homogeneity
and isotropy. Substituting the line element (\ref {eq11}) into
Einstein's equation one gets the interior solution as
\begin{equation}\label{eq16}
\rho_{_{({\rm eff})}}=3\frac{\dot{a}^2}{a^2} = \frac{{\mathcal
M}^{\prime}}{R^{2}R^{\prime}};~~~~ p_{_{({\rm eff})}}
=-\left[\left(\frac{\dot{a}}{{a}}\right)^2+2\frac{\ddot{a}}{a}\right]=
-\frac{\dot{{\mathcal M}}}{R^{2}\dot{R}},
\end{equation}
\begin{equation}\label{eq17}
\dot{R}^{2}=\frac{{\mathcal M}}{R},
\end{equation}
where ${\mathcal M(\tau,r)}$ rises as a free function from the
integration of Einstein's field equation which can be interpreted
physically as the total mass within the collapsing cloud at a
coordinate radius $r$ with ${\mathcal M}\geq0$,~and
$R(\tau,r)=ra(\tau)$ is the physical area radius for the volume
labeled by the comoving coordinate $r$. From equation (\ref{eq14})
and first part of equation (\ref{eq16}), one can solve for the
mass function
\begin{equation}\label{eq18}
{\mathcal M}=\frac{R^3}{3\phi}(\rho_{_{\phi}} + \rho_{m}).
\end{equation}
Using the above equation together with equation
(\ref{eq17}) we arrive at a relation between $\dot{a}$ and the
effective energy density as follows
\begin{equation}\label{eq19}
\dot{a}^2=\frac{a^2}{3\phi}(\rho_{_{\phi}} + \rho_{m}).
\end{equation}
Since we are concerned with a continual collapsing scenario, the
time derivative of the scale factor should be negative
$(\dot{a}<0)$ implying that the physical area radius of the
collapsing volume for constant value of $r$ decreases
monotonically. The singularity arising as the final state of
collapse at $\tau=\tau_{s}$ is given by $a(\tau_{s})=0$. On the
other hand when the scale factor and physical area radius of all
the collapsing shells vanish, the collapsing cloud has reached a
singularity. A point at which the energy density and pressure
blows up, the Kretschmann scalar $ {\mathcal K} =
R^{abcd}R_{abcd}$ diverges and the normal differentiability and
manifold structures break down. In order to solve the field
equations we proceed by substituting for $\rho_{_{({\rm eff})}}$
and $p_{_{({\rm eff})}}$ from equation (\ref{eq16}) into equations
(\ref{eq14}) and (\ref{eq15}) and rewrite them as follows
\begin{equation}\label{eq20}
3\frac{\dot{a}^2}{a^2}=\frac{1}{\alpha m
\Phi}\left[\rho_{_{_{0}}m}a^{-3(1+w)}-3\alpha m
\frac{\dot{a}}{a}\dot{\Phi}+\frac{V(\Phi)}{2}\right],
\end{equation}
\begin{equation}\label{eq21}
-\left[\left(\frac{\dot{a}}{{a}}\right)^2+2\frac{\ddot{a}}{a}\right]=\frac{1}{\alpha
m \Phi}\left[w\rho_{_{_{0}}m}a^{-3(1+w)}+\alpha m
\ddot{\Phi}+2\alpha
m\frac{\dot{a}}{a}\dot{\Phi}-\frac{V(\Phi)}{2}\right],
\end{equation}
where for later convenience we have rescaled the scalar field as
$\phi=\alpha m \Phi$ ($\alpha$ and $m$ are real constants), and by
the virtue of equation (\ref{eq4}) the associated potential to
$f({\mathcal R})=\alpha{\mathcal R}^{m}$ can be fixed as
\begin{equation}\label{eq22}
V(\Phi)=\alpha(m-1)\Phi^{\frac{m}{m-1}}; ~~~~m\neq1.
\end{equation}
Since the scalar field must diverge at the singularity we examine
its behavior by taking the following {\it ansatz} for the scalar
field
\begin{equation}\label{eq23}
\Phi(\tau) = a^{\delta}(\tau),
\end{equation}
where $\delta$ is a constant whose sign decides the divergence
of the scalar field. Substituting the first and second time
derivatives of the scale factor from equations (\ref{eq20}) and
(\ref{eq21}) into equation (\ref{eq7}) together with the use of
equation (\ref{eq22}) one finds
\begin{equation}\label{eq24}
a^{-\left[\delta+3(1+w)\right]}\left\{\frac{2\rho_{_{_{0}}m}(\delta+3w-1)}{3\alpha
m \delta(2+\delta)}
\right\}+a^{\left[\frac{\delta}{m-1}\right]}\left\{\frac{4+\delta(2m-1)-2m}{6m\delta+3m\delta^2}
\right\}=0.
\end{equation}
Matching the powers of scale factor in equation (\ref{eq24}) we
arrive at the following expression for $\delta$
\begin{align}\label{eq25}
\delta=\frac{3(1+w)(1-m)}{m},
\end{align}
whence by substituting the above equation into the pair of square
brackets in equation (\ref{eq24}) one gets the following equation
to be satisfied by $m$
\begin{align}\label{eq26}
\alpha\Big\{3(1+w)+
m\left[-13-9w+m(8+6w)\right]\Big\}+2\rho_{_{_{0}m}}\left[4m-3(1+w)\right]=0.
\end{align}
Solving the above equation, we find $m$ as a function of $\alpha$
and $\rho_{_{_{0}m}}$ for $w$ as
\begin{align}\label{eq27}
m_{\pm}=\frac{13\alpha-8\rho_{_{_{0}m}}\pm\left[73\alpha^2-16\alpha\rho_{_{_{0}m}}+64\rho_{_{_{0}m}}^2\right]^{\frac{1}{2}}}{16\alpha},
\end{align}
\vspace{.4cm}for $w=0$,
\begin{align}\label{eq28}
m_{\pm}=-\frac{-5\alpha+4\rho_{_{_{0}m}}\pm\left[13\alpha^2-16\alpha\rho_{_{_{0}m}}+16\rho_{_{_{0}m}}^2\right]^{\frac{1}{2}}}{6\alpha},
\end{align}
\vspace{.4cm}for $w=-\frac{1}{3}$,
\begin{align}\label{eq29}
m_{\pm}=-\frac{-7\alpha+8\rho_{_{_{0}m}}\pm\left[33\alpha^2-80\alpha\rho_{_{_{0}m}}+64\rho_{_{_{0}m}}^2\right]^{\frac{1}{2}}}{8\alpha},
\end{align}
\vspace{.4cm}for $w=-\frac{2}{3}$, and
\begin{align}\label{eq30}
m_{\pm}=\frac{4\alpha-2\rho_{_{_{0}m}}\pm\sqrt{2}\left[3\alpha^2+2\alpha\rho_{_{_{0}m}}+2\rho_{_{_{0}m}}^2\right]^{\frac{1}{2}}}{5\alpha},
\end{align}
\vspace{.4cm}for $w=\frac{1}{3}$, where $\alpha\neq0$.

\section{Time Behavior Of The Scale Factor and singular epoch}

One would like to study time behavior of the scale factor as the
collapse evolves, considering matter fields. If at time
$\tau=\tau^{*}$ (or equivalently for some $a=a^{*}$) the collapse
begins, then by integrating equation (\ref{eq20})
together with the use of equations (\ref{eq22}) and (\ref{eq23})
near the singularity with respect to time one gets the time
behavior of the scale factor as
\begin{equation}\label{eq31}
a(\tau)=\left[a^{\ast\frac{1}{2}\left(\delta+3(1+w)\right)}-\frac{1}{2}\sqrt{\frac{2\rho_{_{_{0}m}}+\alpha
(m-1)}{6\alpha
m(1+\delta)}}\left(\delta+3(1+w)\right)(\tau-\tau_{\ast})\right]^{\frac{2}{\delta+3(1+w)}},
\end{equation}
and the corresponding singular epoch as
\begin{equation}\label{eq32}
\tau_{s}=2\sqrt{\frac{2\rho_{_{_{0}m}}+\alpha (m-1)}{6\alpha
m(1+\delta)}}\frac{a^{\ast\frac{1}{2}\left(\delta+3(1+w)\right)}}{\delta+3(1+w)}+\tau^{\ast},
\end{equation}
where the time $\tau_{s}$ corresponds to a vanishing scale factor.
Thus the collapse reaches the singularity in a finite proper time.
This result for the scale factor completes the interior solution
within the collapsing cloud, providing us with the required
construction.

\section{The Conditions}

We are now in a position to investigate the nature of singularity
as the endstate of a collapsing scenario. The singularity is
called locally naked if it is only visible to an observer being in
the neighborhood of it (such a singularity is necessarily covered
by a spacetime event horizon) and is called globally naked if
there exists a family of future directed non-spacelike geodesics
reaching to the outside observers in the spacetime and terminating
at the past in the singularity. To investigate the nature of
spacetime singularity arising from the collapse procedure, we
examine here whether such a singularity could be naked, or
necessarily covered within a spacetime event horizon and if so
under what conditions. Since $\alpha$ can be regarded as a free
parameter, we seek for the appropriate values of it satisfying the
following conditions
\begin{itemize}
\item The ratio ${\mathcal M}/R$ stays less than unity till the
singular epoch which means that the singularity has been formed
earlier than the formation of the apparent horizon or the
formation of trapped surfaces have been failed till the singular
epoch. \vspace{.4cm}
\item $\delta<0$ during the gravitational collapse scenario accompanied
by divergence of the scalar field.\vspace{.4cm}
\item For physical reason, weak energy condition, stated as $\rho_{_{({\rm
eff})}}\geq0$ and $\rho_{_{({\rm eff})}}+p_{_{({\rm eff})}}\geq0$
must be satisfied during the dynamical evolution of the
system.\vspace{.4cm}
\item The effective energy density and effective pressure blow up in
the vicinity of the singularity, the latter
being negative during the evolution of the collapse process, since the
absence of trapped surfaces is accompanied by a negative
pressure. In fact the negativeness of the effective pressure implies
that $\dot{{\mathcal M}}<0$, that is, the mass contained in the
collapsing volume with comoving coordinate $r$ keeps decreasing
leading to an outward energy flux during the gravitational collapse scenario.
\vspace{.4cm}
\item Kretschmann scalar diverges at the singular time and then
converges to zero at late times.
\end{itemize}
In order to determine whether the singularity is naked or not, one
needs to investigate the formation of trapped surfaces during the
collapse procedure. These surfaces are defined as compact
two-dimensional (smooth) spacelike surfaces such that both
families of ingoing and outgoing null geodesics orthogonal to them
necessarily converge or the expansion parameter $\Theta$ of the
outgoing future-directed null geodesics is everywhere
negative~\cite{Frolov-Malec}. Consider a congruence of outgoing
radial null geodesics having the tangent vector
$(V^{\tau},V^{r},0,0)$, where $V^{\tau}=d\tau/dk~and~V^{r} =dr/dk$
and $k$ is an affine parameter along the geodesics. For the
spacetime metric (\ref{eq11}), the geodesic expansion parameter
which is defined as the covariant divergence of the vector field
$V^{\nu}$ is given by~\cite{Singh}
\begin{equation}\label{eq33}
\Theta=\nabla_{\nu}V^{\nu}=\frac{2}{r}\left[1-\sqrt{\frac{{\mathcal
M}}{R}}\right]V^{r}.
\end{equation}
If the null geodesics terminate at the singularity in the past
with a definite tangent, then at the singularity we have $\Theta>0$.
If such family of curves do not exist and the event horizon
forms earlier than the singularity, a black hole is
formed. Utilizing equation (\ref{eq33}), we now attempt to study
the formation of trapped surfaces during the
dynamical evolution of the gravitational collapse procedure. We show that
physically, the formation of a black hole or a naked singularity
as the final state for the dynamical evolution is governed by the
rate of collapse and the presence of scalar field. It is
seen that for a specified range of variation of $\alpha$, the
cosmic censorship conjecture may be violated for all cases of
matter considered below. In the following subsections we
consider first the four conditions mentioned in the beginning of
this section and postpone the last one to the next section. We
begin by calculating the ratio ${\mathcal M}/R$ in the general
case which is considered for the four cases of matter,
$w=\left\{0,-\frac{1}{3},-\frac{2}{3},\frac{1}{3}\right\}$
corresponding to dust, cosmic strings, domain walls and radiation,
respectively. By the virtue of equation~(\ref{eq17}) we have
\begin{equation}\label{eq34}
\frac{{\mathcal
M}}{R}=r^2\left[\frac{2\rho_{_{_{0}m}}+\alpha(m-1)}{6\alpha
m(1+\delta)}\right]a^{-(\delta+3w+1)}.
\end{equation}
The weak energy condition which states that the energy density as
measured by any local observer must be non-negative can be written
for any timelike vector $V^{\mu}$ as follows
\begin{equation}\label{eq35}
T_{\mu\nu}V^{\mu}V^{\nu}\geq0,
\end{equation}
whereby one gets the following conditions for the effective energy
density $(\rho_{_{({\rm eff})}}\geq0)$
\begin{equation}\label{eq36}
\rho_{_{({\rm
eff})}}=\left[\frac{2\rho_{_{_{0}m}}+\alpha(m-1)}{2\alpha
m(1+\delta)}\right]a^{-(\delta+3(w+1))}\geq0,
\end{equation}
and the sum of effective energy density and pressure
($\rho_{_{({\rm eff})}}+p_{_{({\rm eff})}}\geq0$) as
\begin{align}\label{eq37}
\rho_{_{({\rm eff})}}+p_{_{({\rm
eff})}}=\left(1+w+\frac{\delta}{3}\right)\left[\frac{2\rho_{_{_{0}m}}+\alpha(m-1)}{2\alpha
m(1+\delta)}\right]a^{-\left(\delta+3(w+1)\right)}\geq0.
\end{align}
Finally for the rate of change of mass function with respect to
time one has
\begin{align}\label{eq38}
\dot{{\mathcal
M}}=r^3(\delta+3w)\left[\frac{2\rho_{_{_{0}m}}+\alpha(m-1)}{6\alpha
m(1+\delta)}\right]a^{-(\delta+3w+1)}\mid\dot{a}\mid,
\end{align}
where the minus sign has been absorbed into $\dot{a}$. At the initial
epoch where $a(\tau^{*})=1$ there should not be any trapping of
light, then by assuming that $r=r_{b}$ is the boundary of the
collapsing volume one may easily see that for a suitable initial
value of energy density the ratio ${\mathcal M}/R$ is less than
unity at the initial time. This fact is in accordance with the
regularity conditions stating that if gravitationally collapsing
massive stars are to be modeled, then the energy density,
pressure, and other physical quantities must be finite and
regular at the initial spacelike hyper-surface from which the
collapse commences. For the case of homogeneous-density collapse
the resulting singularity coincides with the curves $R(\tau_{s}, 0) =
0$ or $R(\tau_{s}, r\neq0) = 0$, corresponding to a central or
non-central singularity, respectively. In the next subsections we
first consider the simpler case of non-central singularity and
investigate formation or otherwise of trapped surfaces for
different values of $w$.

\subsection{Dust ($w=0$)}

For this case of matter we have the following relations (we set
$\rho_{_{_{0}m}}=1$ in the rest of this paper)
\begin{align}\label{eq39}
\left\{\begin{array}{ccccccc}
         \frac{{\mathcal
M}}{R}=r^2\left[\frac{2\rho_{_{_{0}m}}+\alpha(m-1)}{6\alpha
m(1+\delta)}\right]a^{-(\delta+1)}, &  &  &  &  &  &\\
          &  &  &  &  &  &\\
         \rho_{_{({\rm
eff})}}+p_{_{({\rm
eff})}}=\left(1+\frac{\delta}{3}\right)\left[\frac{2\rho_{_{_{0}m}}+\alpha(m-1)}{2\alpha
m(1+\delta)}\right]a^{-(\delta+3)}, &  &  &  &  &
&\left\{\begin{array}{c}
                   \rho_{_{({\rm
eff})}}=\left[\frac{2\rho_{_{_{0}m}}+\alpha(m-1)}{2\alpha
m(1+\delta)}\right]a^{-(\delta+3)}, \\
                      \\
                      \\
                      \\
                     \dot{{\mathcal
M}}=r^3\delta\left[\frac{2\rho_{_{_{0}m}}+\alpha(m-1)}{6\alpha
m(1+\delta)}\right]a^{-(\delta+1)}\mid\dot{a}\mid. \\
                   \end{array}\right.\\
          &  &  &  &  &  &\\
         p_{_{({\rm
eff})}}=\frac{\delta}{3}\left[\frac{2\rho_{_{_{0}m}}+\alpha(m-1)}{2\alpha
m(1+\delta)}\right]a^{-(\delta+3)}, &  &  &  &  &  &\\
       \end{array}\right.
\end{align}
If the interval for $\alpha$ is $-3.9<\alpha<-0.1$ and by
picking out $m_{-}$ in equation (\ref{eq27}), the corresponding
parameter $\delta_{-}$ is always negative, its absolute value
varies between $1<\mid \delta_{-} \mid<3$ and its minimum and
maximum values are $\delta_{-}=-2.72484$ and
$\delta_{-}=-1.00569$, respectively. It is seen that for such
values of $\alpha$ and $\delta_{-}$, the ratio ${\mathcal M}/R$
stays less than unity and then the expansion parameter is always
positive up to the singularity, that is the singularity is formed
earlier than the formation of apparent horizon which is the
boundary of trapped surfaces. The negativeness of effective
pressure for the allowed values of $\alpha$ ensures that
$\dot{{\mathcal M}}$ is negative as collapse proceeds which means
that the mass contained in the collapsing volume keeps waning. Then
there exists an outward energy flux which may be visible to
outside observers (for the case of globally naked singularity)
since the trapped surfaces do not form early enough to cover the
singularity. In addition, the validity of the weak energy
condition can be easily checked by the virtue of expressions
obtained for $\rho_{_{({\rm eff})}}$ and $\rho_{_{({\rm
eff})}}+p_{_{({\rm eff})}}$.

\subsection{Cosmic Strings ($w=-\frac{1}{3}$)}

Cosmic strings are the result of hypothetical 1-dimensional
(spatially) topological defects which may have been constructed during
a symmetry breaking phase transition at the early Universe.
The possibility of their existence was first considered
by Tom Kibble in 1976~\cite{TKible}. A fluid of cosmic strings may
have an effective equation of state,
$p_{_{_m}}=-\frac{1}{3}\rho_{_{_m}}$, so one has the following
relations for this type of matter fluid as
\begin{align}\label{eq40}
\left\{\begin{array}{ccccccc}
         \frac{{\mathcal
M}}{R}=r^2\left[\frac{2\rho_{_{_{0}m}}+\alpha(m-1)}{6\alpha
m(1+\delta)}\right]a^{-\delta}, &  &  &  &  &  &\\
          &  &  &  &  &  &\\
         \rho_{_{({\rm eff})}}+p_{_{({\rm
eff})}}=\left(\frac{\delta+2}{3}\right)\left[\frac{2\rho_{_{_{0}m}}+\alpha(m-1)}{2\alpha
m(1+\delta)}\right]a^{-(\delta+2)}, &  &  &  &  &
&\left\{\begin{array}{c}
                   \rho_{_{({\rm
eff})}}=\left[\frac{2\rho_{_{_{0}m}}+\alpha(m-1)}{2\alpha
m(1+\delta)}\right]a^{-(\delta+2)}, \\
                      \\
                      \\
                      \\
                     \dot{{\mathcal
M}}=r^3(\delta-1)\left[\frac{2\rho_{_{_{0}m}}+\alpha(m-1)}{6\alpha
m(1+\delta)}\right]a^{-\delta}\mid\dot{a}\mid. \\
                   \end{array}\right.\\
          &  &  &  &  &  &\\
         p_{_{({\rm
eff})}}=\left(\frac{\delta-1}{3}\right)\left[\frac{2\rho_{_{_{0}m}}+\alpha(m-1)}{2\alpha
m(1+\delta)}\right]a^{-(\delta+2)}, &  &  &  &  &  &\\
       \end{array}\right.
\end{align}
Choosing $m_{+}$ in equation (\ref{eq28}), it can be seen that the
valid range of change for $\alpha$ is $-100<\alpha<-0.1$ upon
which the absolute value of parameter $\delta_{+}$ is restricted
to vary between $0<\mid \delta_{+}\mid <2$~(in order to prevent
the ratio ${\mathcal M}/R$ to be infinity, $\delta_{+}=-1$ has to
be excluded) and its minimum and maximum values are
$\delta_{+}=-1.86224$ and $\delta_{+}=-0.61558$, respectively. It
should be noted that one can choose the lower limit of $\alpha$ to
be much less than $-100$, but such a choice does not affect
considerably the value of $\delta_{+}$ and its magnitude remains
close to $-0.6$. Taking these values into account, the ratio
${\mathcal M}/R$ stays finite till the singular epoch and causes
the expansion parameter to be positive up to the singularity, and
if no trapped surfaces exist initially then none would form
until the epoch $a(\tau_{s})=0$ which is consistent with the fact
that there exist families of outgoing radial null geodesics
emerging from the singularity. Also the weak energy condition is
satisfied ($\mid \delta_{+}\mid <2$) and the effective pressure is
negative (since $\delta_{+}<0$), consistent with the fact
that time derivative of mass function is negative i.e., the mass
incorporated in the region where the collapse procedure evolves
keeps falling off. As a result, there exists an outward
energy flux during the collapse scenario which may be visible to
external~Universe.

\subsection{Domain Walls ($w = -\frac{2}{3}$)}

Domain walls are two-dimensional objects that form when a discrete
symmetry is spontaneously broken at a phase
transition\cite{Weinberg}. It has been noticed that there exist a
link between domain-walls and cosmologies such as brane
cosmology\cite{Arkani}. The effective equation of state for a
fluid of domain walls may be $p_{_{_m}}=-\frac{2}{3}\rho_{_{_m}}$ and
the mentioned conditions for this case can be written as
\begin{align}\label{eq41}
\left\{\begin{array}{ccccccc}
         \frac{{\mathcal
M}}{R}=r^2\left[\frac{2\rho_{_{_{0}m}}+\alpha(m-1)}{6\alpha
m(1+\delta)}\right]a^{-(\delta-1)}, &  &  &  &  &  &\\
          &  &  &  &  &  &\\
         \rho_{_{({\rm eff})}}+p_{_{({\rm
eff})}}=\left(\frac{\delta+1}{3}\right)\left[\frac{2\rho_{_{_{0}m}}+\alpha(m-1)}{2\alpha
m(1+\delta)}\right]a^{-(\delta+1)}, &  &  &  &  &
&\left\{\begin{array}{c}
                   \rho_{_{({\rm
eff})}}=\left[\frac{2\rho_{_{_{0}m}}+\alpha(m-1)}{2\alpha
m(1+\delta)}\right]a^{-(\delta+1)}, \\
                      \\
                      \\
                      \\
                     \dot{{\mathcal
M}}=r^3(\delta-2)\left[\frac{2\rho_{_{_{0}m}}+\alpha(m-1)}{6\alpha
m(1+\delta)}\right]a^{-(\delta-1)}\mid\dot{a}\mid. \\
                   \end{array}\right.\\
          &  &  &  &  &  &\\
         p_{_{({\rm
eff})}}=\left(\frac{\delta-2}{3}\right)\left[\frac{2\rho_{_{_{0}m}}+\alpha(m-1)}{2\alpha
m(1+\delta)}\right]a^{-(\delta+1)}, &  &  &  &  &  &\\
       \end{array}\right.
\end{align}
As long as $\alpha$ varies in the range $-100<\alpha<-0.1$ (lower
limit of $\alpha$ can be chosen much less than $-100$ but such a
choice has no noticeable influence on the maximum value of
$\delta$), by choosing $m_{+}$ in equation (\ref{eq29}), the
parameter $\delta_{+}$ is negative, its absolute value is less
than unity and its minimum and maximum values are
$\delta_{+}=-0.953501$ and $\delta_{+}=-0.379572$, respectively.
One then may easily see that at initial epoch ($a(\tau^{*}) = 1$),
the regularity condition (there should be no trapped surfaces at
the initial hyper-surface from which the collapse commences) is
satisfied and the ratio of mass function to physical area radius
of the collapsing volume is less than unity during the collapse
procedure denoting that the expansion parameter being positive up
to the singularity. In this case the collapse to a naked
singularity may take place, where the trapped surfaces do not form
early enough or are avoided in the spacetime. Also the mass
contained in the collapsing ball reduces as the time advances due
to the fact that the effective pressure stays negative till the
singular epoch. It is obvious that for such values of $\alpha$ and
$\delta_{+}$ the weak energy condition is satisfied during the
collapse scenario.
\subsection{Radiation ($w=\frac{1}{3}$)}
Finally, for this type of matter we have following relations for
the said conditions
\begin{align}\label{eq42}
\left\{\begin{array}{ccccccc}
         \frac{{\mathcal
M}}{R}=r^2\left[\frac{2\rho_{_{_{0}m}}+\alpha(m-1)}{6\alpha
m(1+\delta)}\right]a^{-(\delta+2)}, &  &  &  &  &  &\\
          &  &  &  &  &  &\\
         \rho_{_{({\rm eff})}}+p_{_{({\rm
eff})}}=\left(\frac{\delta+4}{3}\right)\left[\frac{2\rho_{_{_{0}m}}+\alpha(m-1)}{2\alpha
m(1+\delta)}\right]a^{-(\delta+4)}, &  &  &  &  &
&\left\{\begin{array}{c}
                   \rho_{_{({\rm
eff})}}=\left[\frac{2\rho_{_{_{0}m}}+\alpha(m-1)}{2\alpha
m(1+\delta)}\right]a^{-(\delta+4)}, \\
                      \\
                      \\
                      \\
                     \dot{{\mathcal
M}}=r^3(\delta+1)\left[\frac{2\rho_{_{_{0}m}}+\alpha(m-1)}{6\alpha
m(1+\delta)}\right]a^{-(\delta+2)}\mid\dot{a}\mid. \\
                   \end{array}\right.\\
          &  &  &  &  &  &\\
         p_{_{({\rm
eff})}}=\left(\frac{\delta+1}{3}\right)\left[\frac{2\rho_{_{_{0}m}}+\alpha(m-1)}{2\alpha
m(1+\delta)}\right]a^{-(\delta+4)}, &  &  &  &  &  &\\
       \end{array}\right.
\end{align}
The suitable range of variation of $\alpha$ for $m_{-}$ in
equation (\ref{eq30}) is $-0.66<\alpha<-0.01$ which makes the
corresponding values of $\delta_{-}$ to vary in the range $2<\mid
\delta_{-} \mid<4$ with $\delta_{-}=-2.01007$ and
$\delta_{-}=-3.95037$ are the maximum and minimum values of this
parameter respectively. For such values of $\alpha$ and
$\delta_{-}$ the ratio ${\mathcal M}/R$ stays less than unity as
the collapse procedure ends. Then the expansion parameter remains
positive up to the singularity denoting that the apparent horizon
is failed to form. On the other hand, weak energy condition is
satisfied for these values of $\alpha$ and $\delta_{-}$ and the
effective pressure is negative till the singular epoch. It is
worth noticing that if one chooses the parameter $\alpha$ and the
power of Ricci scalar in equations (\ref{eq27})-(\ref{eq30}) other
than the ones determined in the above subsections, weak energy
condition together with the conditions on effective pressure and
the ratio ${\mathcal M}/R$ would be violated. The central
singularity occurring at $R = 0$, $r = 0$ can be naked if we have
any future-directed outgoing null geodesics terminating in the
past at the singularity. In order to examine the possibility of
existence of such families let us introduce a new variable
$y=r^{\gamma}$ with $\gamma>1$ defined in such away that the ratio
$R'/r^{\gamma-1}$ is a unique finite quantity in the limit
$r\rightarrow0$. Now consider the outgoing radial null geodesic
equation which is given by
\begin{align}\label{radial null}
\frac{d\tau}{dr}=a(\tau).
\end{align}
In terms of variables $y=r^{\gamma}$ and $R$ the above equation
reads
\begin{align}\label{non-cent}
\frac{dR}{dy}=\frac{1}{\gamma
r^{\gamma-1}}\left[\dot{R}\frac{d\tau}{dr}+R'\right],
\end{align}
whence using equation (\ref{eq17}) we have
\begin{align}\label{non-cent}
\frac{dR}{dy}=\frac{R'}{\gamma
y^{\frac{\gamma-1}{\gamma}}}\left[1-\sqrt{\frac{{\mathcal
M}}{R}}\right].
\end{align}
If there exist outgoing radial null geodesics in the past at the
central singularity which occurs at $\tau=\tau_{s}$, then along
such geodesics we have $R\rightarrow0$ as $r\rightarrow0$, or in
terms of the variables $y$ and $R$, the point $y=0$, $R=0$ is a
singularity of the above first order differential equation. For
such a congruence of geodesics $dR/dy$ must be positive up to the
singularity. Then as long as $\alpha$, $\rho_{_{_{0}m}}$, and $m$
satisfy the values which was determined previously, trapped
surfaces would fail to form as the collapse evolves.

\section{nakedness of the singularity}

The final fate of a continual gravitational collapse of a matter
cloud ends in either a black hole or a naked singularity where in
the former there exists an event horizon of spacetime developing
earlier than the formation of the singularity to cover it. Thus
the regions of extreme curvatures and densities are concealed from
the outside observers. The event horizon or surface of a black
hole is defined as the boundary of the spacetime that is causally
connected to future null infinity~\cite{Revisiting EH Finders}, or
in other word the boundary between events from which light rays
emitted inside this boundary surface can not escape to future
infinity while those emitted outside in a suitable direction can.
Both event and apparent horizons coincide in stationary
spacetimes, however this case is not generally true in dynamical
ones. Although the existence of an apparent horizon predicates the
existence of spacetime event horizon, the converse is not always
true and the event horizon may veil the singularity even if
apparent horizon does not emerge on a spatial slice. So far we
have discussed the conditions under which the collapse scenario
ends in a locally naked singularity, that is, the singularity is
visible to an observer being in the neighborhood of it. In this
case, the trajectories coming out of the singularity do not
actually come out to a distant observer but fall back into the
singularity again at a later time without going out of the
boundary of the star. Thus the locally naked singularity could
still be covered by the event horizon and only strong version of
cosmic censorship conjecture is violated but the week form of it
is intact~\cite{TPSinghPSJ}. However if the event horizon is
delayed to form or the singularity forms early enough before the
formation of event horizon, then it would be visible to external
observers and thus a globally naked singularity would born as the
endstate of collapse rather than a black hole. Therefore in such a
situation curvature invariants namely the Kretschmann scalar
should increase near the singularity, diverge at the singular
epoch and then converge to zero at late times~\cite{Kretschmann}.
In previous section we showed that for suitable values of
$\alpha$, trapping of light can be avoided which means that the
apparent horizon is failed to form. But since the failure of
formation of an apparent horizon does not necessarily bode the
absence of an event horizon, we investigate the nakedness of the
singularity in spherically symmetric collapse of a fluid by
considering the behavior of Kretschmann scalar with respect to
time. For the line element (\ref{eq11}) this quantity is given by
\begin{equation}\label{eq50}
{\cal K}\equiv
R^{abcd}R_{abcd}=\frac{12}{a^4}\left[a^2\ddot{a}^2+\dot{a}^4\right].
\end{equation}
By the virtue of equation (\ref{eq31}) one can easily obtain this
quantity as a function of time for
$w=\left\{0,-\frac{1}{3},-\frac{2}{3},\frac{1}{3}\right\}$ and the
results are sketched in Figures 1-4. It is seen that Kretschmann
scalar diverges at singular time and then tends to zero at late
times. In order to better understand the situation one should
resort to critical behavior at the black hole threshold. Such a
behavior was discovered in gravitational collapse of a spherically
symmetric massless scalar field~\cite{14H}, axisymmetric
gravitational waves~\cite{15H}, spherical system of a radiation
fluid~\cite{16H} and spherical system of a perfect fluid obeying
the equation of state $p = w\rho$~\cite{17H}. Consider a type
\textbf{II} critical collapse in which the black hole mass is
scaled as $M_{BH}\propto\mid p-p_{*}\mid^{\gamma}$, where $p$
parameterizes a family of initial data sets evolving through
Einstein's equations, $p_{*}$ is a critical value and $\gamma$ is
a positive constant which is called a critical
exponent~\cite{Kretschmann, Harada-LivingCarsten}. For a
sufficiently large value of $p$ the collapse procedure develops to
a black hole and for a sufficiently small one it evolves to a
dispersion~\cite{Harada-LivingCarsten}. The boundary between these
two regimes is the black hole threshold. Now consider the limit from
supercritical collapse $(p>p_{*})$ to a critical collapse, i.e., $p\rightarrow p_{*}$. In such a limit, the black hole mass tends to
zero and the maximum value of curvature diverges just outside the
event horizon. Since we have arbitrarily strong curvature outside
the event horizon by fine-tuning, the black hole threshold can be
regarded as a globally naked
singularity~\cite{Harada-LivingCarsten}. Let us now consider the
geometry of the exterior spacetime. In order to complete the model
we need to match the interior spacetime of the dynamical collapse
to a suitable exterior geometry. The Schwarzschild solution is a
useful model to describe the spacetime outside stars but the
spacetime outside such a star may be filled with radiated energy
from the star in the form of electromagnetic radiation. The
Schwarzschild solution does not describe this properly as it deals
with an empty spacetime given by $T_{ab}=0$. The spacetime outside
a spherically symmetric star surrounded by radiation emitted from
the star is described by the Vaidya metric\cite{Vaidya} which can
be given in the following form as
\begin{equation}\label{eq51}
ds^{2}_{out}=-\left[1-\frac{2M(r_{v},v)}{r_{v}}\right]dv^{2} -
2dvdr_{v} +r_{v}^2d\Omega^2,
\end{equation}
where $v$ is the retarded null coordinate, $r_{v}$ and
$M(r_{v},v)$ are the Vaidya radius and Vaidya mass, respectively.
In what follows, we use the Isreal-Darmois junction conditions to
match the interior spacetime to a Vaidya exterior geometry at the
boundary hyper-surface $\Sigma$ given by $r = r_{b}$. The
spacetime metric just inside $\Sigma$ is given by
\begin{equation}\label{eq52}
ds^{2}_{in}=-d\tau^2+a^2(\tau)\left[dr^2+r_{b}^2d\Omega^2\right],
\end{equation}
whereby matching the area radius at the boundary one gets
\begin{equation}\label{eq53}
R(r_{b},\tau)=r_{v}(v).
\end{equation}
One then gets the interior and exterior metrics on the
hyper-surface $\Sigma$ as follows
\begin{equation}\label{eq54}
ds^2_{\Sigma in}=-d\tau^2+a^2(\tau)r_{b}^2d\Omega^2,
\end{equation}
\begin{equation}\label{eq55}
ds^2_{\Sigma
out}=-\left[1-\frac{2M(r_{v},v)}{r_{v}}+2\frac{dr_{v}}{dv}\right]dv^2+r_{v}^2d\Omega^2,
\end{equation}
where matching the first fundamental form gives
\begin{equation}\label{eq56}
\left[\frac{dv}{d\tau}\right]_{\Sigma}=\frac{1}{\left[1-\frac{2M(r_{v},v)}{r_{v}}+2\frac{dr_{v}}{dv}\right]^\frac{1}{2}},~~~
(r_{v})_{\Sigma}=r_{b}a(\tau).
\end{equation}
In order to match the second fundamental form (extrinsic
curvature) for interior and exterior spacetimes we need to find
the unit vector field normal to the hyper-surface $\Sigma$. We
then proceed by taking into account the fact that any spacetime
metric can be written locally in the following form as
\begin{equation}\label{56A}
ds^2=-\left(\alpha^2-\beta_{i}\beta^{i}\right)d\tau^2-2\beta_{i}dx^{i}d\tau+h_{ij}dx^{i}dx^{j},
\end{equation}
where $\alpha$, $\beta^{i}$, and $h_{ij}$ are the lapse function,
shift vector, and induced metric, respectively and $i$, $j$ run in
$\{1,2,3\}$. Comparing the above equation with equations
(\ref{eq51}) and (\ref{eq52}) together with using the following
normalization condition for $n^{v}$ and $n^{r_{v}}$
\begin{equation}\label{56B}
n^{v}n_{v} + n^{r_{v}}n_{r_{v}}=1,
\end{equation}
one gets the normal vector fields for the interior and exterior
spacetimes as
\begin{equation}\label{eq57}
n^{a}_{in}=[0,a(\tau)^{-1},0,0],
\end{equation}
\begin{equation}\label{eq58}
n^{v}=-\frac{1}{\left[1-\frac{2M(r_{v},v)}{r_{v}}+2\frac{dr_{v}}{dv}\right]^\frac{1}{2}},~~~~~~~n^{r_{v}}=\frac{1-\frac{2M(r_{v},v)}{r_{v}}+\frac{dr_{v}}{dv}}{\left[1-\frac{2M(r_{v},v)}{r_{v}}+2\frac{dr_{v}}{dv}\right]^\frac{1}{2}}.
\end{equation}
The extrinsic curvature of the hyper-surface $\Sigma$ is defined
as the Lie derivative of the metric tensor with respect to the
normal vector $\textbf{n}$, given by the following relation as
\begin{equation}\label{eq59}
K_{ab}=\frac{1}{2}{\cal
L}_{\textbf{n}}g_{ab}=\frac{1}{2}\left[g_{ab,c}n^{c}+g_{cb}n^{c}_{,a}+g_{ac}n^{c}_{,b}\right],
\end{equation}
whereby the nonzero $\theta$ components of the extrinsic curvature
read
\begin{equation}\label{60}
K^{in}_{\theta\theta}=r_{b}a(\tau),~~~~~~K^{out}_{\theta\theta}=r_{v}\frac{1-\frac{2M(r_{v},v)}{r_{v}}+\frac{dr_{v}}{dv}}{\left[1-\frac{2M(r_{v},v)}{r_{v}}+2\frac{dr_{v}}{dv}\right]^\frac{1}{2}}.
\end{equation}
Setting
$\left[K^{in}_{\theta\theta}-K^{out}_{\theta\theta}\right]_{\Sigma}=0$
on the hyper-surface $\Sigma$, together by using equations
(\ref{eq17}) and (\ref{eq56}) one gets the following relation
between the mass function and Vaidya mass on the boundary as
\begin{equation}\label{eq61}
{\cal M}(\tau,r_{b})=2M(r_{v},v).
\end{equation}
From equation (\ref{eq61}) and equation (\ref{eq18}) it is seen
that the behavior of Vaidya mass is decided by the allowed values
of $\rho_{_{_{0}}m}$, $\alpha$ and $m$ which prompt the
gravitational collapse scenario to end in the formation of a naked
singularity. In order to get a relation describing the rate of
change of Vaidya mass with respect to $r_{v}$ one has to match the
$\tau$ component of the extrinsic curvature on the hyper-surface
$\Sigma$. Setting
$\left[K^{in}_{\tau\tau}-K^{out}_{\tau\tau}\right]=0$ we have
\begin{equation}\label{eq62}
M(r_{v},v)_{,r_{v}}=\frac{{\mathcal M}}{2r_{v}}+r_{b}^2a\ddot{a}.
\end{equation}
Now it can be seen that at the singular time, $\tau=\tau_{s}$ the
ratio $2M(r_{v},v)/r_{v}$ tends to zero. Thus the exterior
spacetime at the singular epoch reads
\begin{equation}\label{62A}
ds^2=-dv^2-2dvdr_{v}+r_{v}^2d\Omega^2,
\end{equation}
which describes a Minkowski spacetime in retarded null
coordinates. Hence, the exterior generalized Vaidya metric at
singular time can be smoothly extended to the Minkowski spacetime
as the collapse completes\cite{PSJ}. The occurrence of a naked
singularity as the final fate of a collapse scenario depends on
the existence of families of non-spacelike trajectories reaching
faraway observers and terminating in the past at the singularity.
In order to show this we begin by equation (\ref{eq56}) and after
using equation (\ref{eq61}) we get
\begin{equation}\label{eq63}
\left[\frac{dv}{d\tau}\right]_{\Sigma}=\frac{1-r_{b}\dot{a}}{1-\frac{{\mathcal
M}(\tau,r_{b})}{r_{v}}}.
\end{equation}
It is seen that imposing the null condition on the Vaidya metric
leads to the same relation as the above. This means that null
geodesics can come out from the singularity and reach distant
observers before it evaporates into the free space. On the other
hand, since for the allowed values of $\alpha$, $\rho_{_{_{0}m}}$
and $m$ formation of trapped surfaces in spacetime is
avoided and from another side the singularity emerges
outside of the event horizon, such a congruence of trajectories
can be detected by the outside observer.
\begin{figure}
\begin{center}
\epsfig{figure=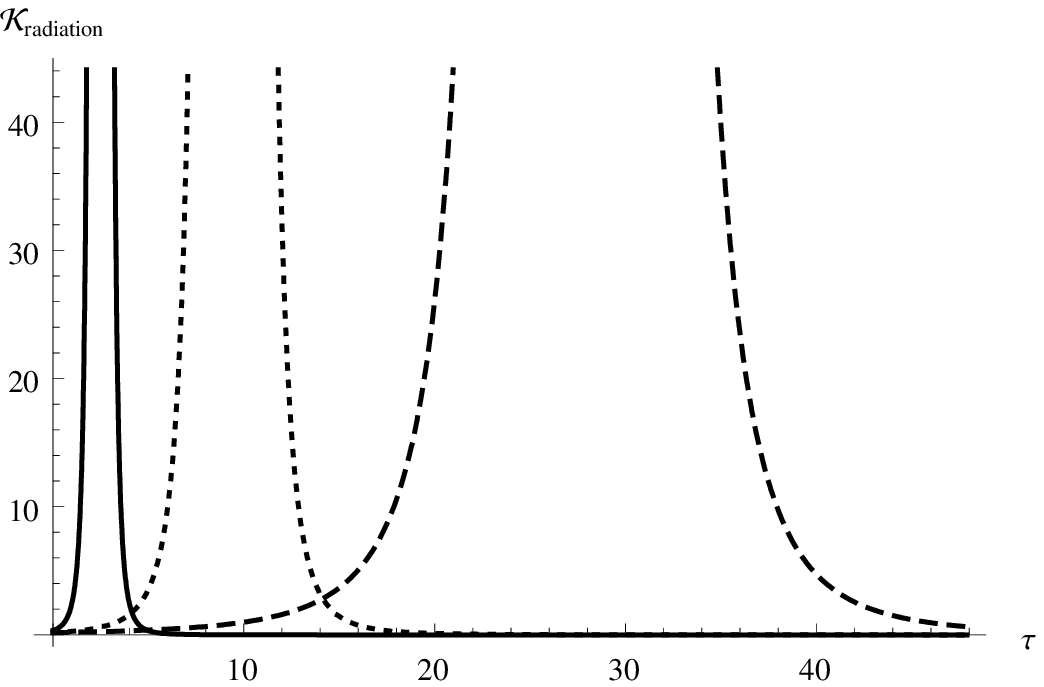,width=7.7cm}\hspace{5mm}
\epsfig{figure=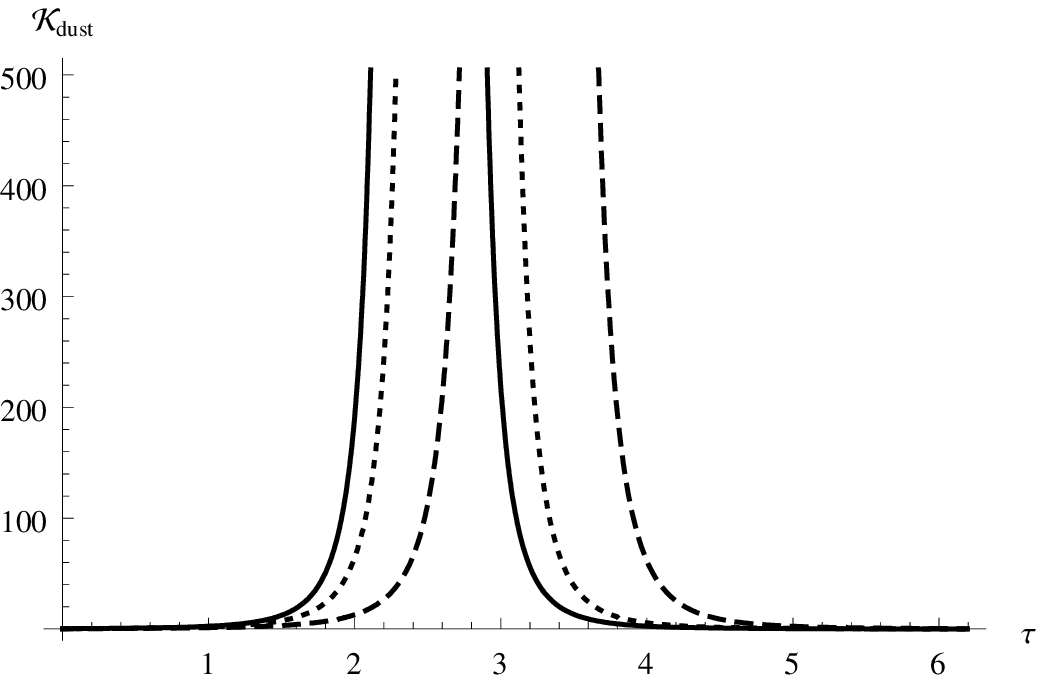,width=7.7cm}\vspace{5mm}
\epsfig{figure=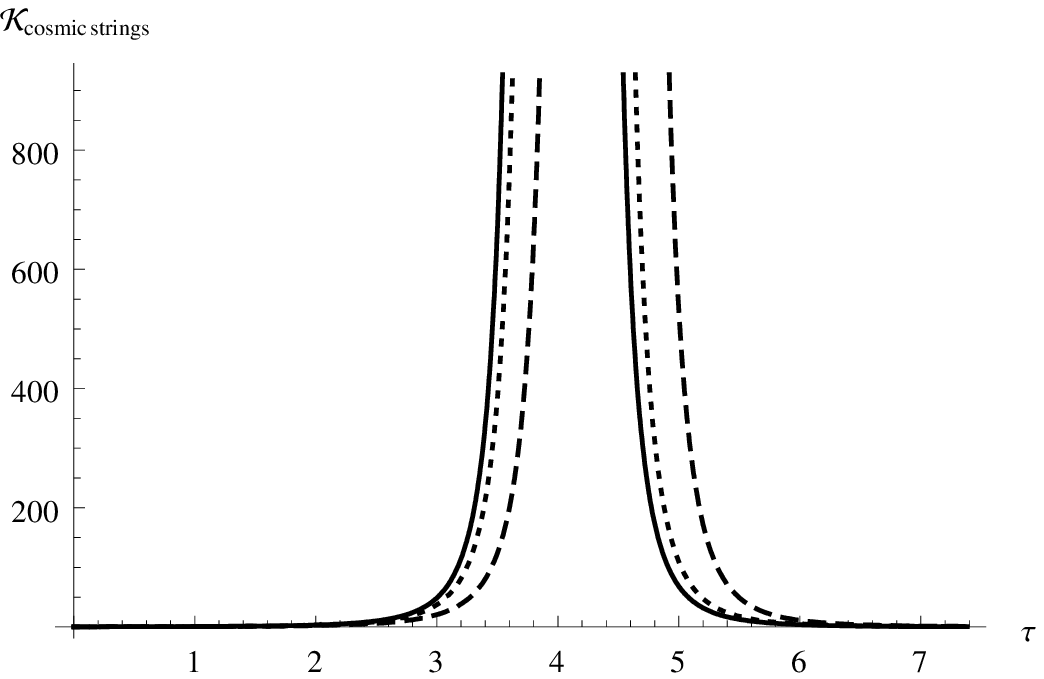,width=7.7cm}\hspace{5mm}
\epsfig{figure=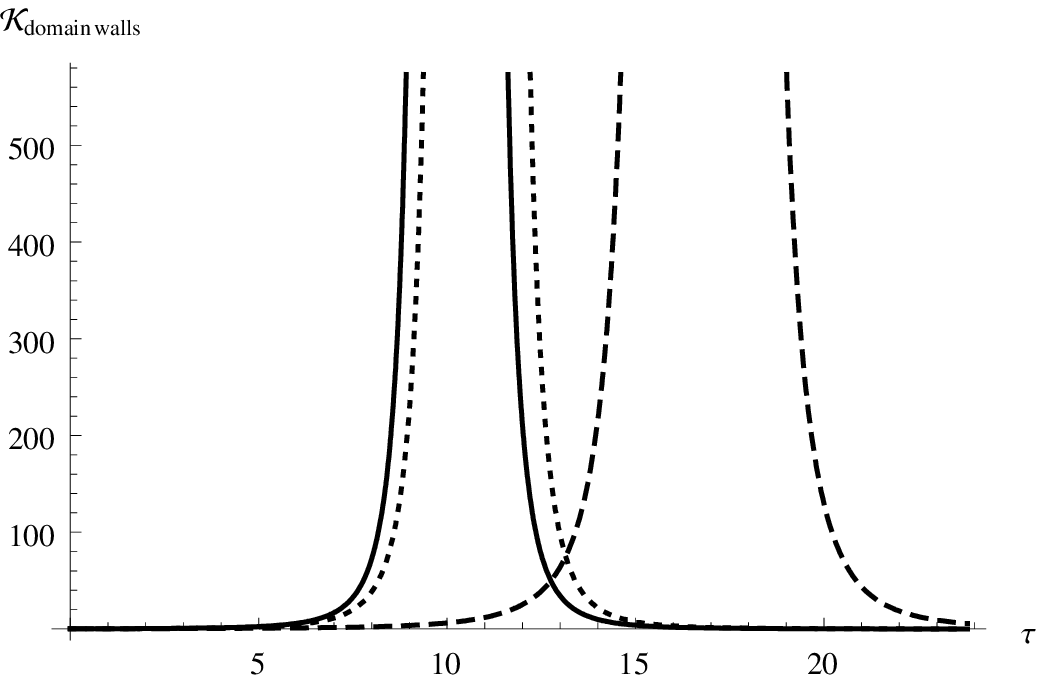,width=7.7cm}
\end{center}
\caption{\footnotesize The behavior of Kretschmann scalar (in
units of $s^{-4}$) as a function of proper time for
$w=\frac{1}{3}$ (upper-left figure) and different values of
$\alpha$ and $\delta_{-}$. $\alpha=-0.65$, $m_{-}=2.02575$ and
$\delta_{-}=-2.02542$, Solid curve, $\alpha=-0.15$,
$m_{-}=5.97355$ and $\delta_{-}=-3.33038$, Dotted curve and
$\alpha=-0.05$, $m_{-}=6.6128$ and $\delta_{-}=-3.75922$, Dashed
curve. For the initial energy density, scale factor, and proper
time we have adopted the values $\rho_{_{0m}}=1$, $a^{\ast}=1$,
and $\tau^{\ast}=0$, respectively. The corresponding singular
epoches are, $\tau_{s}=2.49674$ for Solid curve,
$\tau_{s}=9.44081$ for Dotted curve, and $\tau_{s}=27.8945$ for
Dashed curve.} \caption{\footnotesize The behavior of Kretschmann
scalar (in units of $s^{-4}$) as a function of proper time for
$w=0$ (upper-right figure) and different values of $\alpha$ and
$\delta_{-}$. $\alpha=-3.5$, $m_{-}=1.52406$ and
$\delta_{-}=-1.03175$, Solid curve, $\alpha=-2.5$, $m_{-}=1.60424$
and $\delta_{-}=-1.12996$, Dotted curve and $\alpha=-1.5$,
$m_{-}=1.8076$ and $\delta_{-}=-1.34034$, Dashed curve. For the
initial energy density, scale factor, and proper time we have
adopted the values $\rho_{_{0m}}=1$, $a^{\ast}=1$, and
$\tau^{\ast}=0$, respectively. The corresponding singular epoches
are, $\tau_{s}=2.5083$ for Solid curve, $\tau_{s}=2.70354$ for
Dotted curve, and $\tau_{s}=3.19309$ for Dashed curve.}
\caption{\footnotesize The behavior of Kretschmann scalar (in
units of $s^{-4}$) as a function of proper time for
$w=-\frac{1}{3}$ (lower-left figure) and different values of
$\alpha$ and $\delta_{+}$. $\alpha=-90$, $m_{+}=1.44581$ and
$\delta_{+}=-0.616689$, Solid curve, $\alpha=-30$, $m_{+}=1.46909$
and $\delta_{+}=-0.638609$, Dotted curve and $\alpha=-10$,
$m_{+}=1.54031$ and $\delta_{+}=-0.701562$, Dashed curve. For the
initial energy density, scale factor, and proper time we have
adopted the values $\rho_{_{0m}}=1$, $a^{\ast}=1$, and
$\tau^{\ast}=0$, respectively. The corresponding singular epoches
are, $\tau_{s}=4.04176$ for Solid curve, $\tau_{s}=4.13327$ for
Dotted curve, and $\tau_{s}=4.38509$ for Dashed curve.}
\caption{\footnotesize The behavior of Kretschmann scalar (in
units of $s^{-4}$) as a function of proper time for
$w=-\frac{2}{3}$ (lower-right figure) and different values of
$\alpha$ and $\delta_{+}$. $\alpha=-100$, $m_{+}=1.61179$ and
$\delta_{+}=-0.379572$, Solid curve, $\alpha=-20$, $m_{+}=1.68699$
and $\delta_{+}=-0.407227$, Dotted curve and $\alpha=-2$,
$m_{+}=2.55425$ and $\delta_{+}=-0.608495$, Dashed curve. For the
initial energy density, scale factor, and proper time we have
adopted the values $\rho_{_{0m}}=1$, $a^{\ast}=1$, and
$\tau^{\ast}=0$, respectively. The corresponding singular epoches
are, $\tau_{s}=10.2643$ for Solid curve, $\tau_{s}=10.787$ for
Dotted curve, and $\tau_{s}=16.808$ for Dashed curve.}
\end{figure}
\section{Conclusion and outlook}
One of the physical motivations for discussing naked singularities
is that these objects provide a useful laboratory for quantum
gravity, since in such ultra-strong gravity regions the length and
time scales are comparable to the Planck length and time. In
other words, quantum effects occurring in such super-dense regimes
are no longer covered by the spacetime event horizon and the
chance to observe such effects in the universe is provided. An
example is quantum particle creation due to the formation
of a naked singularity, which has been studied in the
literature~\cite{QPC}. During the past twenty years, cosmic
censorship conjecture has been extensively investigated in
spherical models of gravitational collapse of physically
reasonable matter. The simplest of those which has been
scrutinized in detail and has been shown that both black holes and
naked singularities form from generic initial conditions is
gravitational collapse of a dust fluid. Dwivedi and Joshi in
\cite{Dwivedi-Joshi} and Waugh and Lake in \cite{Waugh-Lake}
showed that a naked strong curvature singularity can be formed in
an inhomogeneous dust collapse and a self-similar one,
respectively. Also some examples of naked singularity formation in
gravitational collapse of a scalar field is given
in~\cite{Christodoulou}.
In this work we have studied the process of gravitational collapse
of a star where the matter fluid obeys the barotropic equation of
state $p=w\rho$, in the context of $f({\mathcal R})$ theories of
gravity. Making use of metric formalism we wrote the action of
$f({\mathcal R})$ gravity as the Brans-Dicke one with vanishing
coupling parameter. Having solved the resulting field equations by
taking the {\it ansatz} (\ref{eq23}) for scalar field we arrived
at the expressions (\ref{eq27})-(\ref{eq30}) for the exponent of
Ricci scalar as a function of $\alpha$ and initial energy density.
In Section \textbf{V} we imposed five conditions on the effective
energy density and pressure, the ratio ${\mathcal M}/R$, parameter
$\delta$, time behavior of the mass function, and Kretschmann
scalar, the validity of which depends on determining
appropriate values of $\alpha$. As long as these conditions are
fulfilled the resulting singularity can be globally naked, i.e.,
ultra-dense regions are no longer covered by a spacetime event
horizon and physical effects are allowed to be shared by the
external Universe. It is worth mentioning that there are future finite-time
singularities in the dark
energy universe coming from modified gravity as well as in
other dark energy theories. Considering $f(\cal{R})$ gravity models
that satisfy cosmological viability conditions (chameleon
mechanism), it is possible to show that finite-time
singularities emerge in several cases. Such singularities
can be classified according to the values of the scale factor $a(t)$,
the density $\rho$, and the pressure $p$ \cite{capo}.
\section{Acknowledgments}
The authors would like to express their sincere thanks to H. R.
Sepangi for useful discussions.

\end{document}